\documentclass[a4paper,12pt]{article}
\usepackage{graphicx}
\usepackage{epsfig}
\usepackage{amssymb}
\usepackage[english]{babel}
\usepackage{amsmath}
\begin{document}
\title{Effect of anisotropy distribution on local nucleation field in bistable ferromagnetic microwires}
\author{Grzegorz Kwiatkowski \footnotemark[1] \footnotemark[2]}
\footnotetext[1]{Science Institute of the University of Iceland, 107 Reykjav\'ik, Iceland}
\footnotetext[2]{Center for Functionalized Magnetic Materials (FunMagMa), Immanuel Kant Baltic Federal University, 236041 Kaliningrad, Russia}

\begin{abstract}
	 Critical parameters defining the local nucleation field in amorphous ferromagnetic microwires with positive magnetostriction are obtained analytically through scaling procedures. Exact value of the nucleation field is obtained numerically as a function of geometric parameters of anisotropy distribution, which is fully taken into account instead of being averaged out. It is established that the value of nucleation field depends predominantly on the steepness of anisotropy change within the boundary between axial and radial domains, while the maximal value of anisotropy inside the wire or an overall average is not relevant.
\end{abstract}

\maketitle

\section{Introduction}
Amorphous ferromagnetic microwires are currently under intensive experimental \cite{zhukova2014fast,zhukova2017effect,baraban2019effect} and theoretical \cite{chiriac1995internal,chiriac2002switching,janutka2015structure} investigation because of their unique and tunable properties such as very strong giant magneto-impedance effect \cite{zhukov2017trends} and bistability with possible applications as components of high performance magnetic and magnetoelastic sensors, actuators \cite{zhukov2009magnetic,zhukov2014magnetic} and smart composites \cite{panina2005stress}. Bistable microwires with positive magnetostriction are of particular interest due to an extremely fast domain wall propagation \cite{zhukov2012manipulation} resulting in quick remagnetisation. There are many experimental works on the dependence of velocity on external magnetic field and mechanical stress \cite{zhukov2012manipulation}, switching field values in various conditions \cite{varga2004switching}, local nucleation field \cite{ipatov2009mechanisms, zhukov2014influence} and shape of the moving domain wall \cite{panina2012domain,ekstrom2010spatial} including the effect of local defects \cite{rodionova2012defects,zhukov2012magnetoelastic}. Yet, there are still many unknowns. In particular, the extremely elongated shape of the moving domain wall needs an explanation and thus more theoretical work is required on the propagation process, since length of the domain wall profile is correlated to propagation speed.

The elongated shape of the moving domain wall \cite{panina2012domain} suggests that there are some important processes occurring near the surface of the wire and indicates an inhomogeneous internal structure of the wire. According to previous theoretical studies \cite{chiriac1995internal}, the anisotropy rapidly changes its magnitude and type from axial to radial near the surface of the microwire. Therefore, taking into account the sharp anisotropy gradient is important for understanding of overall behaviour of microwires. 

While often magnetisation switching in a microwire starts with depinning of a closure domain wall at one of the ends of the wire, which then propagates along the whole wire, the process can be much faster if new domains are nucleated in many places along the wire at the same time effectively cutting the distance each domain wall needs to cover. While such ideas are currently investigated experimentally \cite{ipatov2009mechanisms,ipatov2008local} and some propositions are made to manipulate local nucleation field through induced defects \cite{zhukova2018grading}, such phenomena require more in-depth theoretical studies. In particular, currently available models of domain wall nucleation \cite{aharoni2000introduction} consider only the case of a constant anisotropy which does not fully describe the internal structure of microwires.

	In this work we focus on domain wall nucleation in the middle of a long wire \cite{ipatov2008local,zhukova2014fast} taking into account the rapidly changing anisotropy. Analytic calculations gave us the connection between local nucleation field and material parameters of the wire, while numerical simulations showed the dependence of the nucleation field on the geometry of the wire as well as where exactly a new domain forms.
\section{Methodology}
\begin{figure}
\includegraphics[width=\columnwidth]{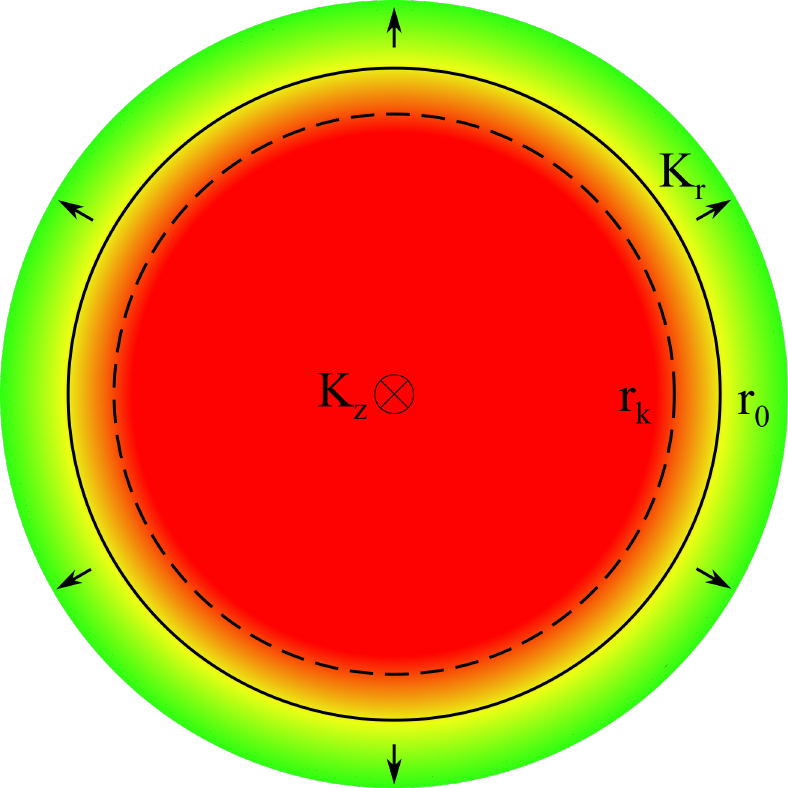}
\caption{Cross-section of the wire. Green (red) colour indicates a region of radial (axial) anisotropy, while yellow is used to show the intermediate layer. Thickness of the intermediate layer has been increased in order to make the picture clearer.}\label{cross}
\end{figure}

\begin{figure}
\includegraphics[width=\columnwidth]{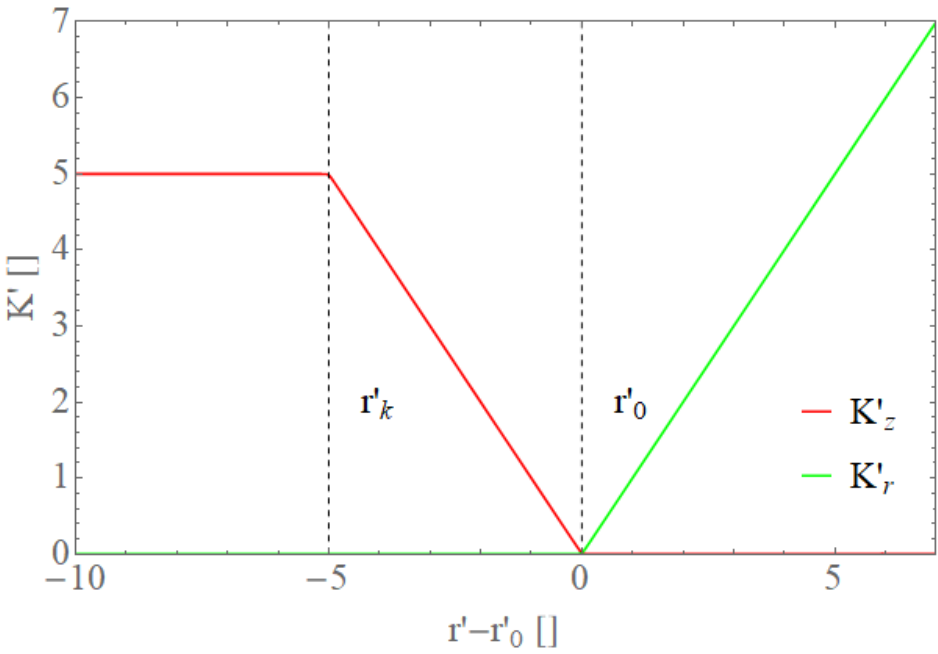}
\caption{Anisotropy distribution in dimensionless reduced variables introduced in Eqs. (\ref{varr})--(\ref{Red}) within the intermediate layer. Axial (radial) anisotropy is shown in red (green).}\label{distr}
\end{figure}

We consider a ferromagnetic microwire described by the Landau-Lifshitz model \cite{landau1935theory}
\begin{equation}\label{eqm}
	 M_s\frac{\partial \overrightarrow{S}}{\partial t}=\gamma\overrightarrow{S}\times\nabla_S E+\alpha\gamma \overrightarrow{S}\times\left(\overrightarrow{S}\times\nabla_S E\right)
\end{equation}
where $M_s$ is the saturation magnetisation, $\overrightarrow{S}$ is the normalised local magnetisation vector, $\gamma$ is the gyromagnetic factor, $\nabla_S$ is a vector variational operator over components of $\overrightarrow{S}$, $\alpha$ is the damping factor and $E$ is the energy density defined as
\begin{eqnarray}
	E & = & \frac{J}{2}\left(\nabla\cdot \overrightarrow{S}\right)^2-K_z \left(\overrightarrow{S}\cdot\hat{z}\right)^2 -K_r\left(\overrightarrow{S}\cdot\hat{r}\right)^2 \nonumber \\ & & -\mu_0 M_s \overrightarrow{H}_{ext}\cdot\overrightarrow{S} -\frac{\mu_0}{2}M_s \overrightarrow{H}_{d}\cdot\overrightarrow{S}
\end{eqnarray}
where $J$ is the exchange energy parameter, $K_z$ and $K_r$ are the coefficients of the easy-axis anisotropy in axial and radial directions respectively (see Fig. \ref{distr} and \ref{cross}), $M_s$ is the saturation magnetisation, $H_{ext}$ and $H_d$ are the external and demagnetazing field respectively.

 Since the coils used to induce local magnetic field in a microwire are significantly longer than the diameter of the wire, the typical nucleation region is long in comparison with the microwire radius. Therefore, we assume a constant external magnetic field as well as a cylindrical symmetry of sought solutions. For the purpose of presenting simulation results we use the angle between the spin vector and the axis of the wire
\begin{eqnarray}
	\theta = \arccos S_z
\end{eqnarray}

	We base our studies on stress distribution obtained by Chiriac and \'{O}v\'{a}ri \cite{chiriac1995internal}, where solidification and cooling stress within glass-coated microwires was calculated using information on internal temperature distribution during the production process. There are later works on stress distribution in microwires, which also include the effect of stress relaxation \cite{antonov2000residual}.

	Stress distribution obtained in \cite{chiriac1995internal} shows that within the core of the wire tensile stress is dominant and slowly rises with increasing radial coordinate up to a point, where it very rapidly drops and switches to strong compressive stress. The angular stress repeats the pattern but always remains lower in value than the axial one. On the other hand the radial stress remains tensile throughout the whole metallic core. Since anisotropy distribution follows the stress pattern $K=\lambda \sigma$ with $\lambda$ being magnetostriction coefficient and $\sigma$ being the magnitude of stress in the leading direction, we can distinguish three distinct areas in the wire (see Fig. \ref{cross}):
\begin{itemize}
\item Surface with very large radial anisotropy.
\item Intermediate layer, where the type and strength of anisotropy changes sharply.
\item Bulk of the wire, where axial anisotropy is dominating and there is little variation in its strength.
\end{itemize}

This pattern is in agreement with known qualities of bistable microwires and their magnetic structure \cite{mohri1990large}.

	Stress and resulting anisotropy distribution is also affected by local defects \cite{zhukov2014influence} as well as intermixing of the metallic core with the glass coating \cite{zhukov2016studies} that can result in a completely different chemical composition of the interfacial layer. Such effects can both influence the radial dependency of anisotropy as well as cause the observed strong variation of local nucleation field along the wire \cite{ipatov2008local, zhukov2014influence}. Nevertheless, the inhomogeneities induced in production or by post-processing \cite{zhukova2018grading,zhukov2017trends} occur on a length scale much larger than the radius of the wire. This means that our approach is still valid for such samples. Inclusion of effects of local defects is beyond the scope of this paper and considered as a topic for further research due to a complex nature of local stress patterns.

	Considering the very high radial anisotropy near the surface perpendicular to the axially-oriented external magnetic field, we assume that magnetic structure in the surface layer does not change significantly during the magnetisation switching, so it will serve as a boundary condition for the intermediate layer. Moreover, the radial domain at the surface is thick enough for us to neglect surface effects in our analysis.

	Since the intermediate layer contains a domain wall between the radial and axial domains (see Fig. \ref{profile}, the case of $H=0$) and the anisotropy is relatively low, we expect the new domain to originate in the intermediate layer and expand into the bulk of the wire. Such a notion is in tune with the empirical estimation of moving domain wall shape obtained by Panina et al. \cite{panina2012domain}, who suggest a propagation from the surface toward centre of the wire rather than along the wire.

 For the purpose of simplification of our model and most importantly extracting key analytical connection between material properties of the wire and local nucleation field we assume the anisotropy coefficients to be linear in radius around the point $r_0$, where the anisotropy changes its type and that both types of anisotropy change at the same rate $d$. The model needs to take into account that the increase in anisotropy toward the centre of the wire stops at some specific radius $r_k$ reaching the bulk value. In particular, we consider (see Fig. \ref{distr})
\begin{eqnarray}
	K_z(r) & = & \frac{d}{4} \left(-r + 2 r_0 - r_k - |r - r_k|\right.\nonumber \\ & & \left. + |r - 2 r_0 + r_k + |r - r_k| |\right) \\
	K_r(r) & = & d \frac{|r-r_0|+r-r_0}{2}
\end{eqnarray}
This form follows the main features of the \textit{ab initio} calculations \cite{chiriac1995internal} without significant deviations especially considering that we focus on a narrow region, where anisotropy changes its character.

	For magnetisation distribution with cylindrical symmetry, the demagnetising field $H_d$ has a simple analytic form (see e.g. \cite{bogdanov1994thermodynamically}):
\begin{equation}
	\overrightarrow{H}_{d}=-M_s \hat{r} \overrightarrow{S}\cdot\hat{r}
\end{equation}
It only introduces a constant shift in radial anisotropy effectively included in further calculations and does not influence qualitative behaviour of the model.

For further analysis it is beneficial to minimise the number of parameters in the equation of motion. In order to perform the reduction, we scale the time and space variables ($t'$ and $r'$ respectively) as well as multiply the whole equation by a constant $C$
\begin{eqnarray}
t'=\frac{\gamma(d^2 J)^{\frac{1}{3}}}{M_s} t \label{varr} \\
r'= \left(\frac{d}{J}\right)^{\frac{1}{3}} r \\
C= (d^2 J)^{-\frac{1}{3}}
\end{eqnarray}
With those procedures we arrive at the dynamical equation containing only four independent and dimensionless parameters two of which ($r'_k$ and $r'_0$) are connected with geometry of the wire and $\alpha$ remains unchanged
\begin{eqnarray}
	H' & = & \mu_0 M_s H_{ext} (d^2 J)^{-\frac{1}{3}} \\
	r'_k & = & \left(\frac{d}{J}\right)^{\frac{1}{3}}r_k \\
	r'_0 & = & \left(\frac{d}{J}\right)^{\frac{1}{3}}r_0 \label{Red}
\end{eqnarray}

Since the damping term does not affect static solutions of the system and their stability, the local nucleation field depends directly on the proportionality coefficient between $H_{ext}$ and $H'$. Therefore, we have already obtained qualitative dependence of critical field value on material parameters, while $H'$ can only depend on $r'_k$ and $r'_0$. Numerical analysis of the system below also provides dependence of nucleation field value on the geometrical parameters of the wire as discussed below. Considering that for a typical microwire the intermediate layer is thin and close to the surface, $r'_0>>r'_0-r'_k$, we use a local Cartesian coordinate system instead of the cylindrical one. This approximation also makes $r'_k-r'_0$ the only relevant parameter of the microwire due to translational invariance. This means that scaled nucleation field $H'_c$ is a function of $r'_k-r'_0$ only
\begin{equation}
	H'_c = H'_c(r'_k-r'_0)
\end{equation}

\section{Results}
Since the scaling procedure (see Eqs. (\ref{varr})--(\ref{Red})) makes the nucleation field $H'_c$ dependent on $r'_k-r'_0$ only, we have
\begin{equation}
	H_c=\frac{(d^2 J)^{\frac{1}{3}}}{\mu_0 M_s} H'_c(r'_k-r'_0)
\end{equation}
where $r'_k-r'_0$ is still dependent on $d$ and $J$ due to scaling -- this will be considered later in the context of numerical results for $H'_c(r'_k-r'_0)$. It is most interesting that the key parameter is the rate of anisotropy change and not the maximal value thereof in the bulk. This means that due to the relationship between anisotropy and stress inside the wire \cite{chiriac1995internal} one can manipulate the nucleation field by modifying the stress gradient inside the wire.

\begin{figure}[h]
\includegraphics[width=\columnwidth]{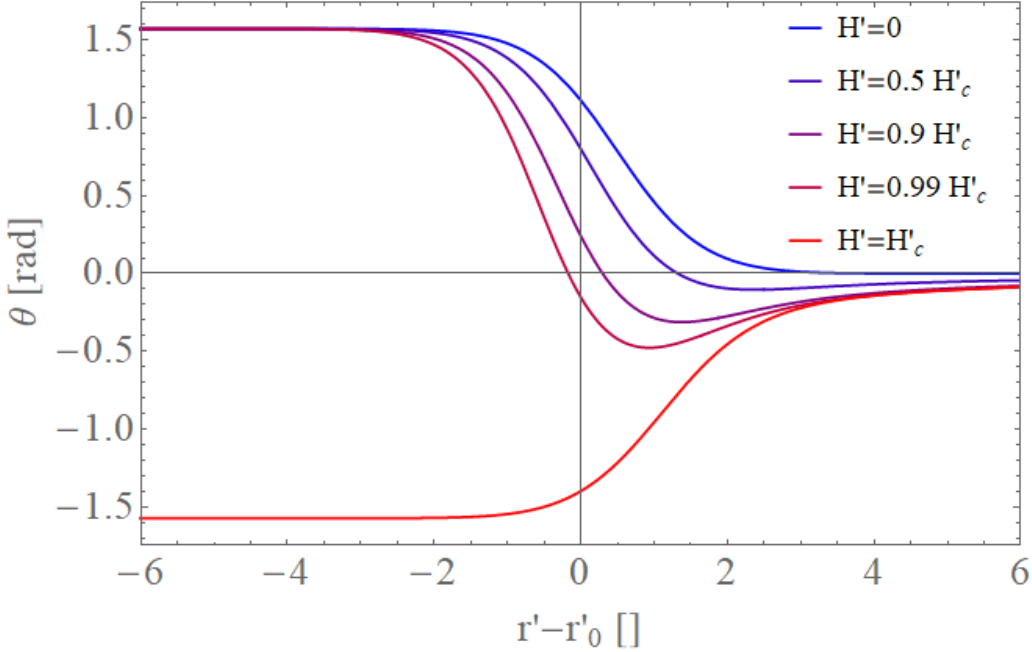}
\caption{Magnetisation profile as a function of dimensionless radius for various intensities of external magnetic field for $r'_k-r'_0=-5$}\label{profile}
\end{figure}

 Equation (\ref{eqm}) is solved numerically using finite difference method. For the purpose of determining the nucleation field for domain nucleation process, we take a static solution for $H'=0$ as an initial value condition and test its stability under magnetic field opposite to the bulk magnetisation direction. For subcritical values of $H'$ the solution deforms, yet stabilises with magnetisation never reaching the same direction as the external magnetic field opposite to the initial magnetisation of the bulk (see Fig. \ref{profile}). For overcritical fields the new domain aligned with the external field forms at the point, where anisotropy is the lowest, and expands unconstrained toward the centre of the wire. Numerical simulations have revealed that a relevant interval for investigation is $r'_k-r'_0>-3$, since the value of nucleation field remains constant below $r'_k-r'_0=-3$.

\begin{figure}
\includegraphics[width=\columnwidth]{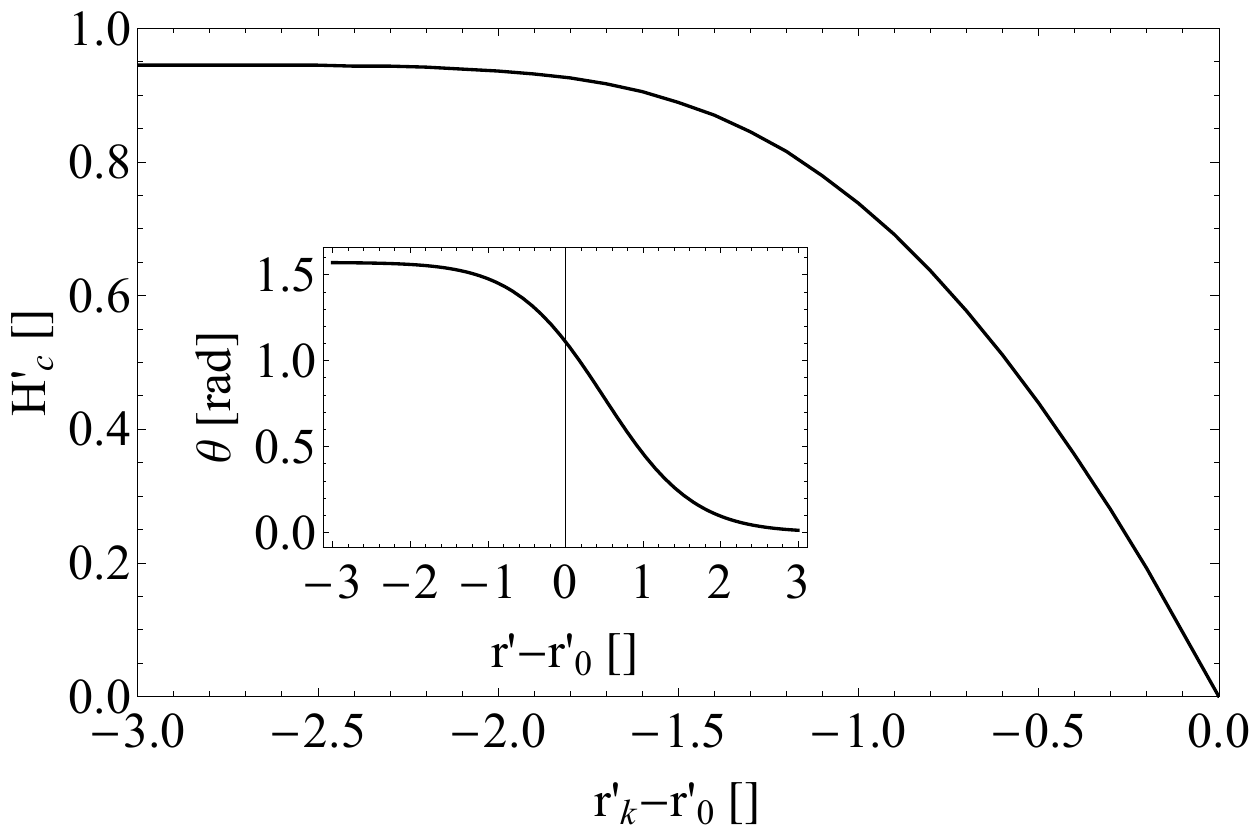}
\caption{Nucleation field as a function of intermediate layer width in dimensionless variables. Inset represents the shape of the domain wall for $r'_k-r'_0=-5$}\label{final}
\end{figure}

	The decrease in nucleation field value, as the width of the intermediate layer becomes smaller, is connected to the size of the initial domain wall between radial and axial domains, but it won't necessarily be observed in typical microwires, since if we take parameter estimations from \cite{chiriac1995internal} and assume the exchange constant for pure iron to be valid for the $Fe_{77.5}Si_{7.5}B_{15}$ compound, then $r'_k-r'_0\approx -5$ which is far in the effectively constant region of the graph (see Fig. \ref{final}). Therefore, the most important result is that there is a specific limit on $H'_c$ value, which means that the nucleation field is not dependent on the maximum anisotropy in the bulk and the anisotropy change rate is the deciding factor. This may seem to contradict results of \cite{ipatov2008local}, where nucleation field is shown to be linearly correlated to average anisotropy (see \cite{aharoni2000introduction} for the derivation of the model). There is, however, an explanation to this. The effective anisotropy is defined by the difference in value between axial and radial stress at a given point. The region, where effective anisotropy changes its type is characterised by equality of the stress components. By adding tensile stress we are shifting the boundary between axial and radial domains toward the surface. Stress is in general not a linear function of radial displacement. In fact, around the intermediate layer we have both $\frac{\partial \sigma_{zz}}{\partial r}<0$ and $\frac{\partial^2 \sigma_{zz}}{\partial r^2}<0$ (see \cite{chiriac1995internal} for stress distribution details). This means that as the boundary between radial and axial domains gets closer to the surface due to external stress, the local anisotropy gradient becomes larger. For this reason we observe a clear correlation between average anisotropy and nucleation field. However, our results can explain significant differences in behaviour between particular wires, which cannot be resolved by the model \cite{aharoni2000introduction}, where only average anisotropy and saturation magnetisation are considered as relevant parameters. A very recent experimental study of domain wall propagation control by stress tuning \cite{baraban2019control} have shown that the correlation between bulk stress and domain wall velocity is in the power of $-0.75$, which considering the connection between local nucleation field and domain wall propagation \cite{rodionova2012defects} suggests that the model proposed in this paper has merit.

The simulations clearly show that the domain nucleation process starts in the intermediate layer and even for subcritical fields the change in magnetisation is clearly visible (see Fig. \ref{profile}). In essence, the boundary between the axial and radial anisotropy regions is magnetically softer than the surrounding area and contains a structural defect in form of a domain wall on which the nucleation of a new domain is easier. This may be the key to understanding of the domain wall propagation process in bistable microwires -- both for the speed and the elongated shape. The general idea for future research would be to consider a movement of a domain wall not along the wire but rather view it as a spreading of domain nucleation process in the intermediate layer and a domain wall movement from the surface toward the centre of the wire. This agrees with correlations between normal ($v_n$) and axial velocity ($v_a$) of a domain wall moving through a microwire $v_a\approx \frac{L}{R} v_n$ (where $L$ is the length of the domain wall along the wire and $R$ is the radius of the wire) observed by Panina et. al. \cite{panina2012domain}.

\section{Conclusions}
	Domain nucleation process clearly starts in the region of the wire, where the anisotropy changes its character and there is a clear analytic relation between the nucleation field value and internal parameters of the wire. Nucleation field depends on the anisotropy change rate, which can be controlled by production conditions, annealing and other processing procedures. Study of domain nucleation process in microwires gives valuable insight into domain wall propagation and gives a direction for further research on the subject.

\section{Acknowledgements}
	This work has been supported by the Icelandic Research Fund (Grant No. 184949-051) and the FunMagMa project of Immanuel Kant Baltic Federal University. I would like to thank Valeria Rodionova and Ksenia Chichay for providing expertise on the state of the art knowledge on amorphous ferromagnetic microwires. I am also grateful to Sergey Leble and Mikhail Vereshchagin for long discussions on the subject and Pavel F. Bessarab for insightful comments on this article.	

\bibliographystyle{unsrt}
\bibliography{refs}
\end{document}